\documentclass[12pt]{article}
\usepackage{amssymb,epsfig,times,graphicx}
\usepackage{amsmath}
\usepackage{pstricks,pst-node,pst-plot}
\usepackage{graphicx}
\numberwithin{equation}{section}

\newtheorem{thm}{Theorem}

\newtheorem{prop}{Proposition}

\newtheorem{exam}{Example}

\newtheorem{rmk}{Remark}

\newcommand{\beq}{\begin{equation}}
\newcommand{\eeq}{\end{equation}}
\newcommand{\de}{\partial}
\def\d{\partial}
\def\f{\frac}
\newcommand{\lm}{\lambda}

\newcommand{\pf}{\noindent{\it Proof \ }}

\newcommand{\epf}{$\quad$\hfill
\raisebox{0.11truecm}{\fbox{}}\par\vskip0.4truecm}
\begin{document}
 
\title{Hamiltonian structure of reductions of the Benney system}

\author{John Gibbons${}^*$, Paolo Lorenzoni${}^{**}$, Andrea Raimondo${}^{*}$\\
\\
{\small * Department of Mathematics, Imperial College}\\
{\small 180 Queen's Gate, London SW7 2AZ, UK}\\
{\small j.gibbons@imperial.ac.uk, a.raimondo@imperial.ac.uk}\\
\\
{\small ** Dipartimento di Matematica e Applicazioni}\\
{\small Universit\`a di Milano-Bicocca}\\
{\small Via Roberto Cozzi 53, I-20125 Milano, Italy}\\
{\small paolo.lorenzoni@unimib.it}}
 
\date{}

\maketitle


\begin{abstract}
We show how to construct the Hamiltonian structures of any
reduction of the Benney chain (dKP) starting from the
family of conformal maps associated to it.
\end{abstract}

\section*{Introduction}
The Benney moment chain \cite{be73}, given by the
equations
\beq\notag
A^k_t=A^{k+1}_x+k A^{k-1}A^0_x, \qquad k=0,1,\dots,
\eeq
with $A^k\!=\!A^k(x,t)$, is the most famous example of a
chain of
hydrodynamic type, which generalizes the classical systems
of
hydrodynamic type in the case when
the dependent variables (and the
equations they have to satisfy) are infinitely many.

A $n-$component reduction of the Benney chain
is a restriction of the infinite dimensional system to a
suitable $n-$dimensional submanifold, that is
$$A^k=A^k(u^1,\dots,u^n), \qquad k=0,1,\dots$$
The reduced systems are systems of hydrodynamic type in
the variables $(u_1,...,u_n)$ that parametrize the
submanifold:
\begin{equation}\notag
u^i_t=v^i_j(u)u^j_x, \qquad i=1,\dots,n.
\end{equation}
Benney reductions were introduced in \cite{gits96}, and
there it was proved that such systems are integrable via
the generalized hodograph transformation \cite{tsa85}.
In particular, this method requires the system to be diagonalizable, that is, there exists a set of coordinates $\lm^1,\dots,\lm^n,$ called \emph{Riemann
invariants}, such that the reduction takes diagonal form:
\beq\notag
\lm^i_t=v^i(\lm)\lm^i_x.
\eeq
The functions $v^i$ are called \emph{characteristic velocities}. 

A more compact description of the Benney chain can be
given by introducing
the formal series
\beq\notag
\lm=p+\sum_{k=0}^{+\infty}\frac{A^k}{p^{k+1}}.
\eeq
In this picture, as follows from \cite{kuma77,kuma78},
the Benney chain can be written as the single equation
$$\lm_t=p\lm_x-A^0_x\lm_p,$$
which is the equation of the second flow of the
dispersionless KP hierarchy. This equation related with
the Benney chain also appears in \cite{lema79}.

Clearly, in the case of a reduction, the coefficients of
this series depend on a finite number of variables
$(u^1,\dots,u^n)$. In this case, the series can be thought
as the asymptotic expansion
  for $p\!\mapsto\!\infty$ of a suitable function
$\lm(p,u^1,\dots,u^n)$  depending piecewise analytically
on the parameter $p$. It turns out \cite{gits96,gits99}
 that such a function satisfies a system of chordal
Loewner equations, describing families of conformal maps
(with respect to $p$) in the complex upper half plane. The
analytic properties of $\lm$ characterize the reduction.
More precisely, in the case of an $n-$reduction the
associated function $\lm$ possess $n$ distinct critical
points on the real axis, these are the characteristic
velocities $v^i$ of the reduced system, and the corresponding
critical values can be chosen as Riemann invariants.

Some examples of such reductions, discussed below, have
known Hamiltonian structures, but the most general result
is far weaker, all such reductions are semi-Hamiltonian
\cite{tsa85,gits96}.

The aim of this paper is to investigate the relations
between the analytic properties of the function
$\lm(p,u^1,\dots,u^n)$ and the Hamiltonian structures of
the associated reduction. As is  well known, such
structures are associated to pseudo-riemannian metrics,
and in particular, local Hamiltonian structures are
associated to flat metrics.

Our approach is general, in the sense that it applies to
all Benney reductions. Consequently, it reveals a unified
structure for the Hamiltonian structure of such reduced
systems.
The main result of the paper provides the
Hamiltonian structures of a Benney reduction directly in terms of
the function $\lm(p,u^1,\dots,u^n)$ and its inverse with respect to $p$, denoted by  $p(\lm,u^1,\dots,u^n)$. The Hamiltonian operator then takes the form
$$\Pi^{ij}=\varphi_i\,\lm^{''}(v^i)\delta^{ij}\frac{d}{dx}+\Gamma^{ij}_k\,\lm^k_x+\frac{1}{2\pi i}\sum_{k=1}^n\int_{C_k}\frac{\frac{\de
p}{\de\lm}\,\,\,\lm^i_x}{(p(\lm)-v^i)^2}\left(\frac{d}{d
x}\right)^{-1}\!\!\!\frac{\frac{\de
p}{\de\lm}\,\,\,\lm^j_x}{(p(\lm)-v^j)^2}\,\varphi_k(\lm)\,d\lm,$$
where
\begin{align*}
\Gamma^{ij}_k\,\lm^k_x&=\frac{\varphi_j\,\lm^i_x-\varphi_i\,\lm^j_x}{(v^i-v^j)^2}\qquad\qquad i\neq j,\\
&\\
\Gamma^{ii}_k\,\lm^k_x&=\varphi_i\,\left(\frac{1}{6}\frac{\lm^{''''}(v^i)}{\lm^{''}(v^i)}
-\frac{1}{4}\frac{\lm^{'''}(v^i)^2}{\lm^{''}(v^i)^2}
\right)\lm^i_x+\frac{1}{2}\varphi^{'}_i\,\lm^i_x-\sum_{k\neq i}\frac{\lm^{''}(v^i)}{\lm^{''}(v^k)}
\frac{\varphi_i\,\lm^k_x}{(v^i-v^k)^2}.
\end{align*}
Here $\varphi_1,\dots,\varphi_n$ are  arbitrary functions of a  single variable, $C_k$ are suitable closed contours on a complex domain, and 
$$\lm^{''}(p)=\frac{\de^2\lm}{\de p^2}(p),\qquad \lm^{'''}(p)=\frac{\de^3\lm}{\de p^3}(p),\,\,\dots$$

\vskip 10mm

The paper is organized as follows. In Section \ref{ssht}
we review the concepts of integrability for diagonalizable
systems of hydrodynamic type and the Hamiltonian formalism
for these systems, both in the local and nonlocal case. In
Section \ref{sbenney} we introduce the Benney chain, its
reductions, and we discuss the properties of these
systems. Section \ref{svlasov} is dedicated to the
representation of Benney reductions in the $\lm$ picture
and to the relations with the Loewner evolution. The study
of the Hamiltonian properties of reductions of Benney is
addressed in Sections \ref{shamc} and \ref{shamv}: in the
former we use a direct approach, starting from the
reduction itself, in the latter we describe these results
from the point of view of the function $\lm$ associated
with the reduction. In the last secion we discuss two
examples   where calculations can be expressed in details.

\section{Systems of hydrodynamic type}\label{ssht}
\setcounter{equation}{0}
 
\subsection{Semi-Hamiltonian systems}
In $(1\!+\!1)$ dimensions, \emph{systems of hydrodynamic type} are quasilinear first order PDE
of the form
\begin{equation}\label{hts}
u^i_t=v^i_j(u)u^j_x, \qquad i=1,\dots,n.
\end{equation}
Here and below sums over repeated indices are assumed
if not otherwise stated. We say that the system (\ref{hts}) is \emph{diagonalizable}
if there exist a set of coordinates $\lm^1,\dots,\lm^n,$ called \emph{Riemann invariants},
such that the matrix $v^i_j(\lm)$ takes diagonal form:
\beq\label{diag}
\lm^i_t=v^i(\lm)\lm^i_x.
\eeq
\noindent
The functions $v^i$ are called \emph{characteristic velocities}. We recall that the Riemann invariants $\lm^i$ are not defined uniquely, but up to a change of coordinates
\beq\label{freedom}
\tilde{\lm}^i=\tilde{\lm}^i(\lm^i).
\eeq
 A diagonal system of
PDEs of hydrodynamic type (\ref{diag})
is called \emph{semi-Hamiltonian} \cite{tsa85}  if the coefficients
$v^i(u)$ satisfy the system of equations
\begin{equation}\label{sh}
\partial_j\left(\frac{\partial_k v^i}{v^i-v^k}\right)=
\partial_k\left(\frac{\partial_j v^i}{v^i-v^j}\right)\hspace{1
cm}\forall i\ne j\ne k\ne i,
\end{equation}
where $\de_i=\frac{\partial}{\partial \lm^i}$. The equations (\ref{sh}) are the integrability conditions both for
the system
\begin{equation}
\label{SYM} \frac{\de_j w^i}{w^i-w^j}=\frac{\de_j v^i}{v^i-v^j},
\end{equation}
which provides  the characteristic velocities  of the symmetries
\begin{equation*}
u^i_{\tau}=w^i(u)u^i_x\hspace{1 cm}i=1,...,n
\end{equation*}
of (\ref{hts}), and for the system
\begin{equation*}
(v^i-v^j)\de_i\de_j H=\de_i v^j\de_j H-\de_j v^i\de_i H,
\end{equation*}
which provides   the densities $H$  of conservation laws of
(\ref{hts}). The properties of being diagonalizable and semi-Hamiltonian
imply the integrability of the system:
\\
\begin{thm}\cite{tsa85}(Generalized hodograph transformation)
 
\noindent
Let
\beq\label{diag2}
\lm^i_t=v^i\lm^i_x
\eeq
be a diagonal semi-Hamiltonian system of hydrodynamic type, and let $(w^1,\dots,w^N)$ be
the characteristic velocities of one of its symmetries. Then, the functions $(\lm^1(x,t),\dots,\lm^N(x,t))$
determined by the system of equations
\beq\label{hodo}
w^i=v^i\,x+t, \qquad i=1,\dots,N,
\eeq
satisfy (\ref{diag2}). Moreover, every smooth solution of this system is locally obtainable in this way.
\end{thm}
 
\subsection{Hamiltonian formalism}\label{hamfor}
 
A class of Hamiltonian formalisms for systems of
 hydrodynamic type \eqref{hts} was introduced by Dubrovin and Novikov
in \cite{duno83,duno84}. They considered local Hamiltonian operators of the form
\begin{equation}\label{PB}
P^{ij}=g^{ij}(u)\frac{d}{dx}-g^{is}\Gamma^{j}_{sk}(u)u^k_x
\end{equation}
and the associated Poisson brackets
\begin{equation}
\label{PBHT} \{F,G\}:=\int\frac{\delta F}{\delta
u^i}P^{ij}\frac{\delta G}{\delta u^j}dx
\end{equation}
where $F=\int g(u)dx$ and $G=\int g(u)dx$ are functionals not
depending on the derivatives $u_x,$ $u_{xx},$...
\begin{thm}\cite{duno83}
If $\det{g^{ij}}\ne 0$, then the formula (\ref{PBHT}) with (\ref{PB})
 defines a Poisson bracket if and only if
 the tensor $g^{ij}$ defines a flat pseudo-riemannian metric and the
coefficients
 $\Gamma^j_{sk}$ are the Christoffel symbols of the associated
Levi-Civita connection.
\end{thm}
Non-local extensions of the bracket (\ref{PBHT}), related to metrics
of constant curvature, were considered by Ferapontov and Mokhov in
\cite{fermok90}. Further generalizations were considered by Ferapontov in \cite{fe91}, where he introduced the nonlocal differential operator
\beq\label{ferop}
P^{ij}=g^{ij}\frac{d}{dx}-g^{is}\Gamma^j_{sk}u^k_x+\sum_{\alpha}\epsilon_{\alpha}\left(w^\alpha\right)^i_ku^k_x
\left(\frac{d}{dx}\right)^{\!-1}\!\!\!\left(w^\alpha\right)^j_hu^h_x\,,\qquad\epsilon_{\alpha}=\pm 1.
\eeq
The index $\alpha$ can take values on a finite or infinite -- even continuous -- set.
\begin{thm}
If $\det{g^{ij}}\ne 0$, then the formula (\ref{PBHT}) with (\ref{PB})
 defines a Poisson bracket if and only if
 the tensor $g^{ij}$ defines a pseudo-riemannian metric, the
coefficients $\Gamma^j_{sk}$ are the Christoffel symbols of the associated
Levi-Civita connection $\nabla$, and the affinors $w^\alpha$ satisfy the conditions
\begin{gather*}
\left[w^\alpha,w^\beta\right]=0,\\
\\
g_{ik}(w^{\alpha})^k_j=g_{jk}(w^{\alpha})^k_i,\\
\\
\nabla_k(w^{\alpha})^i_j=\nabla_j(w^{\alpha})^i_k,\\
\\
R^{ij}_{kh}=\sum_\alpha \left\{\left(w^\alpha\right)^i_k\left(w^\alpha\right)^j_h
-\left(w^\alpha\right)^j_k\left(w^\alpha\right)^i_h\right\},
\end{gather*}
where $R^{ij}_{kh}=g^{is}R^{j}_{skh}$ are the components of the Riemann curvature
tensor of the metric $g$.
\end{thm}
In the case of zero curvature, operator (\ref{ferop}) reduces to (\ref{PBHT}).
Let us focus our attention on semi-Hamiltonian systems.
In \cite{fe91} Ferapontov conjectured that
 any diagonalizable semi-Hamiltonian system is always Hamiltonian with respect to
suitable,
 possibly non local, Hamiltonian operators.
 Moreover he proposed the following construction to define
 such  Hamiltonian operators:
\begin{enumerate}

\item Consider a diagonal system \eqref{diag}. Find the general solution of the system
\begin{equation}
\label{meq} \de_j\ln{\sqrt{g_{ii}}}=\frac{\de_j v^i}{v^j-v^i},
\end{equation}
which is compatible for a semi-Hamiltonian system, and compute the
curvature tensor of the metric $g$.
 
\item If the non vanishing components of the curvature tensor can be written
in terms of solutions $w_{\alpha}^i$ of the linear system (\ref{SYM}):
\begin{equation}\label{exp}
R^{ij}_{ij}=\sum_{\alpha}\epsilon_\alpha
w^i_{\alpha}w^j_{\alpha},\hspace{1 cm}\epsilon_{\alpha}=\pm 1,
\end{equation}
 
\noindent
then it turns out that the system (\ref{hts}) is Hamiltonian with respect
to the Hamiltonian operator
\begin{equation}\label{NLPB}
P^{ij}=g^{ii}\delta^{ij}\frac{d}{dx}-g^{ii}\Gamma^{j}_{ik}(u)u^k_x
+\sum_{\alpha}\epsilon_\alpha
w^i_{\alpha}u^i_x\left(\frac{d}{dx}\right)^{-1}\!\!w^j_{\alpha}u^j_x,
\end{equation}
which is the form of (\ref{ferop}) in case of diagonal matrices.
\end{enumerate}

\section{Benney reductions}\label{sbenney}
 
A natural generalization of $n-$component systems of hydrodynamic type (\ref{hts}) can be obtained by allowing the number of equations and variables to be infinite. These systems are known as \emph{hydrodynamic chains}, and the best known example is the Benney chain \cite{be73}: 

 \beq\label{benney}
A^k_t=A^{k+1}_x+k A^{k-1}A^0_x, \qquad k=0,1,\dots.
\eeq
In this setting, the variables $A^n$ are usually called moments. In \cite{be73} Benney proved that this system admits an infinite series of conserved quantities,
whose densities are polynomial in the moments. The first few of them are
$$H^0=A^0, \quad H^1=A^1, \quad H^2=\frac{1}{2}A^2+\frac{1}{2}\left(A^0\right)^2\dots$$
 
A \emph{$n-$component reduction} of the Benney chain (\ref{benney})
is a restriction of the infinite dimensional system to a suitable $n-$dimensional submanifold in the space of the moments, that is:
\beq\label{conred}
A^k=A^k\left(u^1,\dots,u^n\right), \qquad k=0,1,\dots
\eeq
where $u^i=u^i(x,t)$ are the new dependent variables. These are regarded as coordinates on the submanifold specified by (\ref{conred}), and all the equations of the chain have to be satisfied on this submanifold.
In addition, we require the $x-$derivatives $u^i_x$ to be linearly independent \footnote{If this constraint is relaxed, solutions such as described in \cite{komame02} may be obtained.}, in the
sense that
\beq\label{strongred}
\sum_{i=1}^n\alpha_i(u^1,\dots,u^n) \,u^i_x=0\quad\Rightarrow\quad\alpha_i(u^1,\dots,u^n)=0,
\qquad \forall\,i.
\eeq

Thus, the infinite dimensional system reduces to a system with finitely many dependent variables (\ref{hts}). It was shown in \cite{gits96} that all Benney
reductions are diagonalizable and possess the semi-Hamiltonian
property, hence they are integrable via the generalized hodograph method.
On the other hand, we may consider whether a diagonal system of hydrodynamic
type
\beq\label{diag1}
\lm^i_t=v^i(\lm)\lm^i_x, \qquad i=1,\dots,n.
\eeq
is a reduction of Benney (note that we do not impose the semi-Hamiltonian
condition). A direct substitution in the chain (\ref{benney}) leads, after collecting the $\lm^i_x$ and making use of \eqref{strongred}, to the system
\beq\label{benneyred}
v^i\de_iA^k=\de_iA^{k+1}+kA^{k-1}\de_iA^0, \qquad i=1,\dots,n,
\eeq
where $\de_i A^0=\frac{\de A^0}{\de \lm^i}$. The consistency conditions $$\de_j\de_i A^{k+1}\!=\!\de_i\de_j A^{k+1},\qquad i\neq j,\quad k=0,1,\dots$$ reduce to the $\frac{3}{2}n(n-1)$ equations
\begin{subequations}\label{gt}
\begin{align}
\de_iv^j&=\frac{\de_iA^0}{v^i-v^j}\label{gt1}\\
&\hspace{3cm} i\neq j,\notag\\
\de^2_{ij}A^0&=\frac{2\de_iA^0\de_jA^0}{(v^i-v^j)^2}\label{gt2}
\end{align}
\end{subequations}
which are called the \emph{Gibbons-Tsarev system}. It has been shown that this system is in involution, hence it characterizes a $n$-component reduction of Benney. Moreover, if a solution of (\ref{gt}) is known, all the higher
moments can be found, making recursive use of conditions (\ref{benneyred}).

\begin{thm}\cite{gits96}
A diagonal system of hydrodynamic type (\ref{diag1})  is a reduction of the Benney moment chain (\ref{benney}) if and only if there
exist a function $A^0(\lm^1,\dots,\lm^n)$ such that $A^0$ and the $v^1,\dots,v^n$
of the system satisfy the Gibbons-Tsarev system
(\ref{gt}). In this case, system (\ref{diag1})  is automatically semi-Hamiltonian.
\end{thm}

It was noticed in \cite{gits96} that a generic solution of the Gibbons-Tsarev system depends on $n$ arbitrary functions of one variable. Essentially, this
is due to the fact that in the system (\ref{gt}) the derivatives
\beq\label{diagterms}
\de_i v^i, \qquad \de^2_{ii}A^0
\eeq
are not specified. This leads to a freedom of $2n$ functions of a single variable, which reduces to $n$ allowing for the freedom of reparametrization
(\ref{freedom}) in the definition of Riemann invariants. Thus, for any fixed integer $n$, the Benney moment chain possesses infinitely many integrable $n$-component reductions, parametrized by $n$ arbitrary functions of one variable.

In the next sections we will see how the knowledge of the $\lq$diagonal'
terms (\ref{diagterms}) plays an important role in determining
the Hamiltonian structure of a Benney reduction. If these terms are specified,
the Gibbons-Tsarev system becomes a system of pfaffian type, and a generic solution depends on $n$ arbitrary constants.

\begin{exam}\label{exzak1}
The $2-$component Zakharov reduction \cite{za80}, is obtained by
imposing on the moments the constraints
$$A^k=u^1 \left(u^2\right)^k, \qquad \,k=0,1,\dots,$$
where $(u^1,u^2)$ are the new dependent variables. The resulting classical shallow water wave system, first solved by Riemann, is known to be the dispersionless limit of the $2-$component vector NLS equation. Under the change of dependent coordinates
$$\lm^1=u^2+2\sqrt{u^1} \qquad \lm^2=u^2-2\sqrt{u^1},$$
the system takes the diagonal form (\ref{diag}), with velocities
$$v^1= \frac{3}{4}\lm^1+\frac{1}{4}\lm^2\qquad v^2=\frac{1}{4}\lm^1+\frac{3}{4}\lm^2.$$
\noindent
It is easy to check that these velocities satisfy the Gibbons-Tsarev system
with
$$A^0=\frac{(\lm^1-\lm^2)^2}{16}.$$ 
\end{exam}

\section{The $\lm$ picture and chordal Loewner equations}\label{svlasov}
\subsection{Reductions in the $\lm$ picture}
A more compact description of the Benney chain can be given by introducing
\cite{kuma77}
a formal series
\beq\label{genfun}
\lm(p,x,t)=p+\sum_{k=0}^{\infty}\frac{A^k(x,t)}{p^{k+1}}.
\eeq
It is well known that the moments satisfy the Benney chain
(\ref{benney}) if and only if $\lm$ satisfies
\beq\label{vlasov}
\lm_t=p\lm_x-A^0_x\lm_p=\left\{\lm\, ,\,\frac{1}{2}\left(\lm^2\right)_{\geq 0}\right\},
\eeq
where $\left(\,\,\,\right)_{\geq 0}$ denotes the polynomial part of the argument, and $\{\cdot , \cdot\}$ is the canonical Poisson bracket on the $(x,p)$space.
Equation \ref{vlasov} corresponds to the Lax equation of the second flow of the dispersionless KP hierarchy.
 
\begin{rmk}
If we introduce the inverse of the series $\lm$ with respect to $p$, and
denote it as
$$p(\lm)=\lm+\sum_{k=0}^\infty\frac{H_k}{\lm^{k+1}},$$
then it is easy to check that the following equation holds
$$p_t=\de_x\left(\frac{1}{2}p^2+A^0\right).$$
Equivalently, its coefficients satisfy
$$H^k_t=\de_x\left(H^{k+1}_x-\frac{1}{2}\sum_{i=0}^{k-1}H^iH^{k-1-i}\right),$$
which is the Benney chain written in conservation law form using the coordinate set $H^n$. It is easy to show that every $H^k$ is polynomial in the
moments $A^0,\dots,A^k$.
\end{rmk}
 
The use of the formal series (\ref{genfun}) is to be understood as an algebraic model for describing the underlying integrable system  in a more compact way . However, to describe the system in more detail we must
impose more structure on $\lm$. Following \cite{gits99,giyu00}, rather than considering a formal series in the parameter $p$, we instead consider a piecewise analytic function for the variable $p$. In particular, we let $\lm_+$ be an analytic  function defined on $Im(p)>0$, and $\lm_-$ an analytic function  on $Im(p)<0$. We
also require the normalization
\beq\label{hydronor}
\lm_\pm = p+O\left(\frac{1}{p}\right), \qquad p\mapsto\infty.
\eeq
Let us define, on the real axis, the jump function
$$f(p,x,t)=\frac{1}{2\pi i}\left(\lm_-(p,x,t)-\lm_+(p,x,t)\right),$$
and suppose $f$ is a function of real $p$ which is Holder continuous and satisfying
the conditions
$$\int_{-\infty}^{+\infty} p^n f dp<\infty, \qquad n=0,1,\dots.$$
Then, using Plemelj's formula for boundary values of analytic functions, we may take
$$\lm_\pm(p)=p-\pi\int_{-\infty}^{+\infty} \frac{f(p')}{p-p'}dp'\mp i\pi
f(p).$$
What we obtained is that, with hypotheses above, the functions $\lm_+$ and $\lm_-$ are Borel sums of the series (\ref{genfun}) in the upper and lower half plane respectively. On the other hand, $\lm_\pm$ will have, at $p\mapsto \infty$, the formal asymptotic series
(\ref{genfun}), where
$$A^n(x,t)=\int_{-\infty}^{+\infty} p^n f(p,x,t) dp.$$
Thus, to any solution of Benney's equations we can associate a pair of functions $\lm_\pm(p;x,t)$.
In particular, a real valued $f$ leads to real valued moments. In this case,
using the Schwarz reflection principle, we can restrict our attention to the function $\lm_+$; this is the case studied in \cite{giko94,giyu00,bagi03,bagi04,bagi06}.

On the other hand, it will be useful below to consider the analytic continuation of $\lm_+$ into the lower half plane, in the neighborhood of specified points in the real axis. Such a continuation may or may not coincide with $\lm_-$, the Schwarz reflection of $\lm_+$. In particular important examples such a continuation may be developed consistently, giving the structure of a Riemann surface. 
 
\begin{rmk}\label{normrmk}
Other normalizations, more general than \eqref{hydronor} are allowed, based on the fact that for any differentiable function $\varphi$ of a single variable, the composed function $\varphi(\lambda_+)$ remains a solution of \eqref{vlasov}, the associated reduction being the same. In concrete examples, it is sometimes more convenient to make use a different normalisation.
\end{rmk}
 
Let us consider now the relations between solutions of (\ref{vlasov}) and
Benney reduction. In this case, we have that $\lm_+$
is associated with a $n$ component reduction if and only if it depends on the variables $x,t$ via $n$ independent functions.
As any reduction is diagonalizable, it is not restrictive to take as these
variables the Riemann invariants. Thus, we have
\beq\label{con1}
\lm_+(p,x,t)=\lm_+(p,\lm^1(x,t),\dots,\lm^n(x,t)),
\eeq
with
\beq\label{con2}
\lm^i_t=v^i\lm^i_x.
\eeq
Remarkably, the characteristic velocities of
the reduction turn out to be the critical points of the function $\lm_+$
associated with it. More precisely, we have
 
\begin{thm}
Let $\lm_+$, solution of (\ref{vlasov}), satisfy conditions (\ref{con1})
with (\ref{con2}). Let us denote
$$\varphi^i(\lm^1,\dots,\lm^n)=\lm_+(v^i,\lm^1,\dots,\lm^n), \qquad i=1\dots n,$$
and suppose that the $\rho^i$ are not constant functions. Then, the velocities $v^i$ satisfy
$$\frac{\de\lm_+}{\de p}(v^i)=0, \qquad i=1\dots n,$$
and the corresponding critical values $\varphi^i$ can be chosen as Riemann invariants for the system (\ref{con2}).
\end{thm}
 
\pf
Considering equation (\ref{vlasov}) at $p=v^i$, we obtain the system of $n$
equations
\beq\notag
\varphi^i_t=v^i\varphi^i_x-A^0_x\,\,\frac{\de\lm_+}{\de p}(v^i).
\eeq
As $\lm^1,\dots,\lm^n$ can be chosen as coordinates, by the chain rule we
get
\beq\notag
\sum_{j=0}^n\frac{\de\varphi^i}{\de\lm^j}\,\lm^j_t=v^i\sum_{j=0}^n\frac{\de\varphi^i}{\de\lm^j}\,\lm^j_x-\frac{\de\lm_+}{\de p}(v^i)\sum_{j=0}^n \frac{\de A^0}{\de\lm^j}\,\lm^j_x,
\eeq
and this, after substituting (\ref{con2}) into it, is equivalent to
\beq\label{cdim}
\frac{\de\varphi^i}{\de\lm^j}\left(v^j-v^i\right)+\frac{\de\lm_+}{\de p}(v^i)\,\frac{\de A^0}{\de\lm^j}=0 \qquad i,j=1\dots,n,
\eeq
due to the independence of the $\lm^j_x$. Particularly, for $i=j$ the system above reduces to
\beq\label{ccrit}
\frac{\de\lm_+}{\de p}(v^i)\,\frac{\de A^0}{\de\lm^i}=0.
\eeq
Further, if $A^0$ does not depend on $\lm^i$, the function $\lm_+$ is also independent of the same $\lm^i$. In this case, the associated system (\ref{con2}) reduces to a $n\!-\!1$ reduction. On the other hand, if the system is a proper $n-$component reduction then $\de_iA^0\neq 0$ and the characteristic velocities are critical points for $\lm_+$. Substituting back (\ref{ccrit}) into (\ref{cdim}), we obtain $\varphi^i=\varphi^i\left(\lm^i\right)$. Thus, if the critical values $\varphi^i$ are not constant functions, it is possible to choose them as Riemann invariants.
 
\epf
 
\noindent
The converse of the Theorem above is also true: if $\lm_+$ is a solution of (\ref{vlasov}) satisfying
\beq
\lm_+(p,x,t)=\lm_+(p,\lm^1(x,t),\dots,\lm^n(x,t)),
\eeq
and with $n$ distinct critical points $v^1,\dots,v^m$, then by evaluating equation (\ref{vlasov}) at $p=v^i$ we obtain the diagonal system
$$\varphi^i_t=v^i\,\varphi^i_x,$$
where $\varphi^i=\lm_+(v^i)$. Thus, critical points are characteristic velocities. Moreover, the existence of a function $\lm$ associated with a reduction selects a natural set of Riemann invariants, the critical values of $\lm$. Unless otherwise stated, these are the coordinates we will consider below.

\begin{rmk}\label{extraz}
It might happen that the function $\lm_+$ possesses $m$ critical
points, with $m>n$. This is the case, for instance, in Remark \ref{normrmk}, where critical points of the function $\varphi$ have to be added. Then, substituting the critical points into \eqref{vlasov} we obtain an $m$ component diagonal system. However, in this case we have that $m-n$ of the critical values have trivial dynamics for they are independent of $x,t$. Consequently, the $m$ component system reduces to an $n$ component one.
\end{rmk}

\begin{exam}\label{lmzak}
Consider $u^i=u^i(x,t)$, $i=1,2$. The function
\beq
\lambda_+=p+\frac{u^1}{p-u^2},
\eeq
rational in $p,$ satisfies equation (\ref{vlasov}) if and only if $u^1$, $u^2$ satisfy the $2$ component Zakharov reduction of Example \ref{exzak1}.
\end{exam}

\subsection{Reductions and Loewner equations}
It was shown in \cite{giko89,giyu00} that the solution of the initial value
problem of an $n$ reduction is given by a Inverse Scattering Transform procedure
(which leads to a particular form of Tsarev's generalized hodograph formula
(\ref{hodo})),
provided that
$$\frac{\de \lm_+}{\de p}(p)\neq 0,\qquad Im(p)>0.$$
It is thus necessary that $\lm_+(p)$ be an \emph{univalent conformal map} from the
upper half plane to some image region. In \cite{gits96,gits99} it was proved that
these conformal maps have to be solutions of a system of so called chordal Loewner equations.
In fact, if a solution of equation (\ref{vlasov}) is associated with a $n-$component reduction of Benney, then conditions (\ref{con1}) holds. Substituting into equation (\ref{vlasov}), if $v^i$ are the characteristic velocities
associated with the reduction, we obtain
\begin{equation}
\sum_{i=1}^N \left((v^i-p)\,\frac{\partial \lambda_+}{\partial \lm^i}
+\frac{\partial A^0}{\partial \lm^i} \frac{\de\lambda_+}{\de p}\right) \lm^i_x = 0.
\end{equation}
As the $\lm^i_x$ are independent, then it follows that
\begin{equation}\label{loewner}
\frac{\partial \lambda_+}{\partial \lm^i}= \frac{\de_i
A^0}{p-v^i} \frac{\partial \lambda_+}{\partial p},
\qquad i=1,\dots,n.
\end{equation}
This is a system of $n$ chordal Loewner equations (see for example \cite{du83}). When the function
$\lm_+$ is chosen with the normalization (\ref{hydronor}), this system describes the evolution of families of univalent conformal
maps from the upper complex half plane to the upper half plane
with $n$ slits, when the end points of the slits are allowed to move
along prescribed mutually non intersecting Jordan arcs. Using the implicit function theorem it is
possible to show that the inverse function $p$ satisfies an analogous system
\beq\label{loewp}
\frac{\partial p}{\partial \lm^i}= -\frac{\de_i A^0}{p-v^i},
\qquad i=1,\dots,n.
\eeq
\begin{center}
\begin{figure}[ht]
\centering
\begin{pspicture}(0,-0,5)(13.5,3.5)
\psline(0,0)(1,0)
\psline[linewidth=0.5mm]{*-*}(1,0)(3.5,0)
\psline(3.5,0)(4.5,0)
\psline[linewidth=0.5mm]{*-*}(4.5,0)(7,0)
\psline(7,0)(7.75,0)
\psline[linestyle=dashed](7.75,0)(9.25,0)
\psline(9.25,0)(10,0)
\psline[linewidth=0.5mm]{*-*}(10,0)(12.5,0)
\psline(12.5,0)(13.5,0)
\rput(2.6,0.4){$v^1$}
\rput(2.25,0){$\times$}
\rput(6.1,0.4){$v^2$}
\rput(5.75,0){$\times$}
\rput(11.6,0.4){$v^n$}
\rput(11.25,0){$\times$}
\rput(12.5,1.3){\psframebox{\,$p$\,}}
\end{pspicture}
\centering
\begin{pspicture}(-0.75,-0.5)(12.75,3.5)
\psline(-0.75,0)(7.5,0)
\psline[linestyle=dashed](7.5,0)(9,0)
\psline(9,0)(12.75,0)
\pscurve[linewidth=0.5mm]{*-}(2,0)(3.5,1)(1.5,2)
\pscurve[linewidth=0.5mm]{*-}(10,0)(8.5,1.3)(9.5,2)
\pscurve[linewidth=0.5mm]{*-}(5,0)(6,1)(4.5,1.5)(5.2,2.5)
\rput(1.5,2){$\times$}
\rput(2.3,2.4){$\lm_+(v^1)$}
\rput(5.2,2.5){$\times$}
\rput(6.1,2.7){$\lm_+(v^2)$}
\rput(9.5,2){$\times$}
\rput(10.4,2.2){$\lm_+(v^n)$}
\rput(11.7,1.3){\psframebox{ $ \lm_+$}}
\end{pspicture}
\caption{ $n-$slit Loewner evolution on the upper half plane.}
\end{figure}
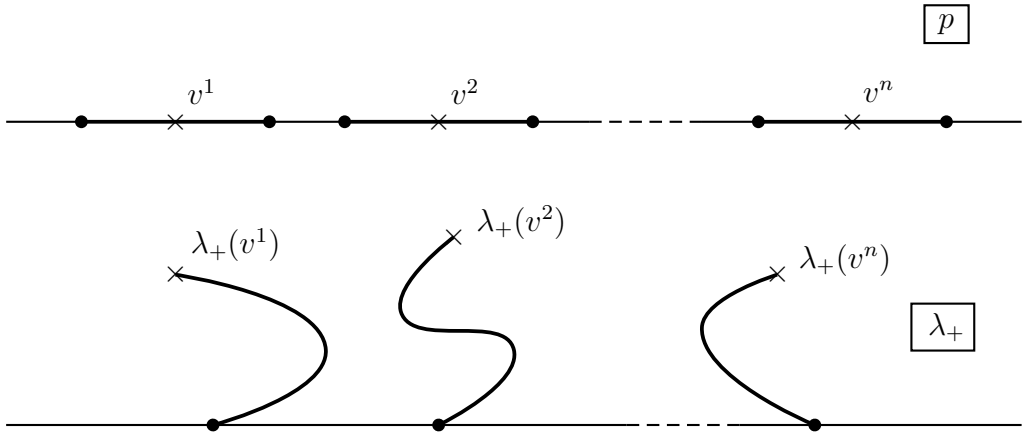
 
\end{center}
For $n>1$, the consistency conditions of (\ref{loewner}) (or (\ref{loewp}) equivalently) turn out to be the Gibbons-Tsarev system. On the other hand,
and more generally, we can consider a set of $n$ Loewner equations,
\beq
\frac{\de \lm_+}{\de\lm^i}=\frac{a_i}{p-b^i}\,\frac{\de\lm_+}{\de p}, \qquad i=1,\dots,n,
\eeq
for arbitrary functions $a_i$, $b^i$. The consistency conditions
$$\frac{\de^2\lm_+}{\de\lm^i\de\lm^j}=\frac{\de^2\lm_+}{\de\lm^j\de\lm^i}$$
are then equivalent to the set of equations
\begin{gather}
\de_i a_j=\de_j a_i\\
\de_i a_j =\frac{2 a_i a_j}{(b^i-b^j)^2}\label{fakegt1}\\
\de_i b^j=\frac{a_i}{b^i-b^j},\label{fakegt2}
\end{gather}
where $i\neq j$. The first of these equations implies locally the existence of a function
$A^0(\lm^1,\dots, \lm^n)$ such that
$$a_i=\de_i A^0.$$
Consequently, equations (\ref{fakegt1}), (\ref{fakegt2}) become the Gibbons-Tsarev
system (\ref{gt}), with $b^i=v^i$. So, to any solution of a system of $n$ chordal Loewner equations there corresponds a $n$-component reduction of the Benney chain.
 
\begin{exam}\label{dbouss}
The dispersionless Boussinesq reduction, which is a $2-$component Gelfand-Dikii reduction, is given by
\begin{align*}
A^0_t&=A^1_x\\
A^1_t&=-A^0A^0_x,
\end{align*}
can be described in the $\lm$ picture using the polynomial function
$$\lm_+=p^3+3A^0p+3A^1.$$
The characteristic velocities are
$$v^1=-\sqrt{-A^0},\qquad v^2=\sqrt{-A^0},$$
and the Riemann invariants are given by
$$\lm^1=\lm_+(v^1)=3A^1+2(-A^0)^{\frac{3}{2}}, \qquad \lm^2=\lm_+(v^2)=3A^1-2(-A^0)^{\frac{3}{2}}.$$
After the renormalization
$$\tilde{\lm}(\lm_+)=\sqrt[3]{\lm_+}=\sqrt[3]{p^3+3A^0p+3A^1}$$
we obtain a family of Schwarz-Christoffel maps as in Figure \ref{figbou}. It is easy to verify that the critical points of the function $\tilde{\lm}(\lm_+)(p)$ are the same as $\lm_+(p)$, while the corresponding new Riemann invariants are $\tilde{\lm}^i=\sqrt[3]{\lm^i}$.
\begin{center}
\begin{figure}[ht]\label{figbou}
\centering
\begin{pspicture}(0,-0,5)(13.5,3.5)
\psline(0,0)(4,0)
\psline[linewidth=0.5mm]{*-*}(2.75,0)(6.75,0)
\psline[linewidth=0.5mm]{*-*}(6.75,0)(10.75,0)
\psline(9.25,0)(13.5,0)
\rput(5.1,0.4){$v^1$}
\rput(4.75,0){$\times$}
\rput(9.1,0.4){$v^2$}
\rput(8.75,0){$\times$}
\rput(12.5,1.3){\psframebox{\,$p$\,}}
\end{pspicture}
\centering
\begin{pspicture}(0,-0.5)(13.5,3.5)
\psline(0,0)(13.5,0)
\psline(9,0)(12.75,0)
\psline[linewidth=0.5mm]{*-}(6.75,0)(5,2.5)
\rput(5,2.5){$\times$}
\rput(5.5,2.7){$\tilde{\lm}^1$}
\psline[linewidth=0.5mm]{*-}(6.75,0)(8.2,1.7)
\rput(8.2,1.7){$\times$}
\rput(8.7,1.9){$\tilde{\lm}^2$}
\rput(12.7,1.3){\psframebox{ $ \lm_+$}}
\end{pspicture}
\caption{The dispersionless Boussinesq reduction.}
\end{figure}
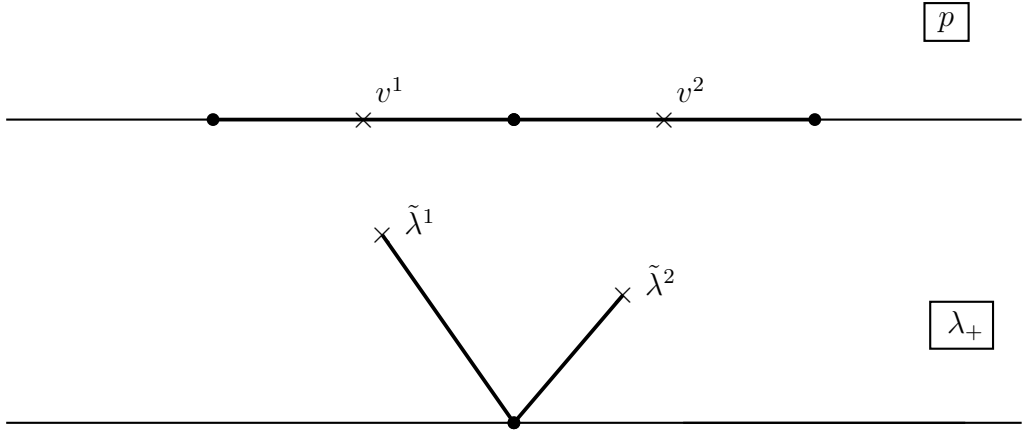
\end{center}
\end{exam}

As a consequence of the Loewner system \eqref{loewner} satisfied by a Benney reduction, it follows immediately that the critical points of $\lm_+(p)$ are simple. Indeed, taking the limit of the $i-$th equation of the system, for $p\to v^i$ gives
\beq\label{2lma}
1=\frac{\de^2\lm_+}{\de p^2}(v^i)\,\,\de_i A^0,
\eeq 
where we used the identity 
$$\frac{\de \lm}{\de\lm^i}|_{\,p=v^i}=\frac{d\lm^i}{d\lm^i}-\frac{\de\lm}{\de p}|_{\,p=v^i}\frac{\de v^i}{\de \lm^i}=1.$$ 
Thus, 
$$\frac{\de^2\lm_+}{\de p^2}(v^i)\neq 0,$$
hence the $v^i$ are simple.

Suppose now that $\lm_+$ admits an analytic continuation in some neighborhood of $v^i$. Henceforth, to simplify the notations,  the subscript $+$ will be dropped from $\lm_+$, and we will denote both the analytic function $\lm_+$ and its analytic continuation simply by $\lm$. Moreover, we will write
$$\lm^{'}(p)=\frac{\de\lm}{\de p}(p),\quad \lm^{''}(p)=\frac{\de^2\lm}{\de p^2}(p), \quad \dots$$
Then, the function $p(\lm)$ has the series development near $\lm=\lm^i,$
\beq\label{pexp}
p(\lm)=v^i+\frac{\sqrt{2}}{\sqrt{\lm^{''}(v^i)}}\sqrt{\lm-\lm^i}+O\left(\lm-\lm^i\right),
\eeq
which becomes a Taylor expansion in the complex local parameter $t=\sqrt{\lm-\lm^i}$. Furthermore, we have
\begin{align}\label{oneoverlm}
\frac{1}{\lambda^{'}(p)}=&\frac{1}{\lm^{''}(v^k)}\frac{1}{p-v^k}-\frac{1}{2}\frac{\lm^{'''}(v^k)}{\lm^{''}(v^k)^2}\notag\\
&\\
&+\left(\frac{1}{4}\frac{\lm^{'''}(v^k)^2}{\lm^{''}(v^k)^3}-\frac{1}{6}\frac{\lm^{''''}(v^k)}{\lm^{''}(v^k)^2}\right)
(p-v^k)+O\left((p-v^k)^2\right).\notag
\end{align}
This expansion will be useful in Section \ref{shamv}. Finally, we introduce two sets of contours, in the $p$ and $\lambda$ plane respectively, that we will need later for describing the Hamiltonian structure of the reductions. We define $\Gamma_i$ as a closed and sufficiently small contour in the $p-$plane around $v^i$, and  $C_i$ as the image of $\Gamma_i$ according to the analytical continuation of $\lm$. Thus, $\Gamma_i$ and $C_i$ are well defined; in particular, it follows from expansion \eqref{pexp} that $\lm_i$ -- the tip of the slit -- is a square root branch point for $p(\lm)$, hence $C_i$ encircles it twice.

\subsection{Symmetries of the Benney reductions}
A well-known method \cite{giko89} of obtaining a countable
set of symmetries of the Benney reduction
  is based on the Lax representation of the dKP hierarchy,
$$\lm_{t_n}=\left\{\lm\,,
\,h_n\right\}=\left(h_n\right)_p\lm_x-\left(h_n\right)_x\lm_p,
\qquad
n=1,2,\dots$$
where $h_n=\frac{1}{n}\left(\lm^n\right)_{\geq 0}$. We assume, unless otherwise stated, that $\lm$ is normalized as in \eqref{hydronor}. If the solution $\lm$ possesses $n$ critical
values $(v^1,\dots,v^n)$ , the hierarchy can be reduced to
$$\lm^i_{t_n}=w_n^i\lm^i_x, \qquad i=1,\dots,n,\qquad
n=1,2,\dots,$$
with
\beq\label{symben}
w_n^i=\left(\frac{\de h_n}{\de p}\right)_{|\,p=v^i}.
\eeq
These are, by construction, components of the symmetries
of the reductions, the first few of them being
$$w^i_1=1,\quad w^i_2=v^i, \quad
w^i_3=(v^i)^2+A^0\,\,\dots \qquad i=1,\dots,n.$$
 
Further, the above characteristic velocities can be obtained as the coefficients of the expansion at $\lambda=\infty$
  of the generating functions
\beq\label{gfs}
W^i(\lambda)=\frac{1}{p(\lambda)-v^i}.
\eeq
 
\begin{prop}
The functions $W^i(\lm)$ are solutions of the linear
system
\beq\label{eqgenf}
\frac{\de_j W^i(\lm)}{W^j(\lm)-W^i(\lm)}=\frac{\de_j
v^i}{v^j-v^i}.
\eeq
Moreover, expanding $W^i(\lm)$ at $\lambda\!=\!\infty$ we get
\beq\label{expsym}
W^i(\lm)=\sum_{n=1}^\infty\,\frac{w^i_n}{\lm^{n}},
\eeq
where the coefficients of the series are the symmetries $w_n^i$ given in \eqref{symben}.
\end{prop}
\begin{rmk}
The first condition \eqref{eqgenf} holds in any normalization, but the expansion \eqref{expsym} assumes the normalization \eqref{hydronor}.
\end{rmk} 
\vskip 5mm
\pf In order to prove \eqref{eqgenf}, knowing that
$$\frac{\de p}{\de\lm^i}=\frac{\de_i A^0}{v^i-p},$$
we can write
\begin{gather*}
\de_j W^i(\lm)=\de_j\left(\frac{1}{p-v^i}\right)=\frac{-1}{(p-v^i)^2}\left(\frac{\de p}{\de\lm^j}-\frac{\de v^i}{\de\lm^j}\right)\\
=-\frac{\de_j A^0}{(p-v^i)^2}\left(\frac{1}{v^j-p}-\frac{1}{v^j-v^i}\right)\\
=\frac{\de_jA^0}{(p-v^i)(p-v^j)(v^j-v^i)}.
\end{gather*}
On the other hand
\begin{gather*}
W^j(\lm)-W^i(\lm)=\frac{(v^j-v^i)}{(p-v^i)(p-v^j)}.
\end{gather*}
and so
$$\frac{\de_j W^i(\lm)}{W^j(\lm)-W^i(\lm)}=\frac{\de_j
A^0}{(v^j-v^i)^2}=\frac{\de_j v^i}{v^j-v^i}.$$
 
\vskip 10 mm
\noindent
In order to prove \eqref{expsym}, chosen the normalization \eqref{hydronor}, we have
\begin{equation*}
w^i_n=\frac{1}{n}\,\lim_{p\to v_i}\frac{d}{dp}\left[(p-v_i)^2
\frac{(\lambda^n)_{+}}{(p-v_i)^2}\right]=\frac{1}{n}\,\underset{p=v_i}{\rm{res}}\left(\frac{(\lambda^n)_{+}}{(p-v_i)^2}\,dp\right).
\end{equation*}
The function $\frac{(\lambda^n)_{+}}{(p-v_i)^2}$ has poles
only at $p=v_i$ and $p=\infty$
  and therefore
\begin{eqnarray*}
&&\underset{p=v_i}{\rm{res}}\left(\frac{(\lambda^n)_{+}}{(p-v_i)^2}\,dp\right)=
-\underset{p=\infty}{\rm{res}}\left(\frac{(\lambda^n)_{+}}{(p-v_i)^2}\,dp\right)=
-\underset{p=\infty}{\rm{res}}\left(\frac{(\lambda^n)}{(p-v_i)^2}\,dp\right)
\end{eqnarray*}
where the last identity is due to
\begin{equation*}
\underset{p=\infty}{\rm{res}}\left(\frac{(\lambda^n)_{-}}{(p-v_i)^2}\,dp\right)=0.
\end{equation*}
Changing variable we obtain
\begin{align*}
-\underset{p=\infty}{\rm{res}}\left(\frac{(\lambda^n)}{(p-v_i)^2}\,dp\right)=&
-\frac{1}{2\pi
i}\int_{\Gamma_\infty}\frac{\lambda^n\,\,\frac{dp}{d\lambda}}{(p(\lm)-v_i)^2}\,d\lambda\\
=&\,\,\,\frac{n}{2\pi
i}\int_{\Gamma_\infty}\frac{\lambda^{n-1}}{p(\lm)-v_i}\,d\lambda,
\end{align*}
where $\Gamma_\infty$ is a sufficiently small contour around
$p=\infty$. Thus,
$$w^i_n=\int_{\Gamma_\infty} W^i(\lm)\lm^{n-1}\,d\lm,$$
varying $n$ we obtain the coefficients of the expansion \eqref{expsym}.
\begin{flushright}
$\Box$
\end{flushright}

\begin{rmk}It will be useful below to consider, as a generating function of the symmetries, 
\beq\label{symgenfun}
w^i(\lambda)=\frac{\frac{\de
p}{\de\lambda}}{(p(\lambda)-v^i)^2}=-\sum_{n=1}^\infty\,\frac{n\,\,w^i_n}{\lm^{n+1}},
\eeq
which is nothing but the $\lambda-$ derivative of \eqref{gfs}.
\end{rmk}

\noindent 
We finally observe that, due to linearity of \eqref{eqgenf},  the functions
$$z^i=\sum_{k=1}^n\int_{C_k}w^i(\lambda)\varphi_k(\lambda)d\lambda$$
still satisfy the system for the symmetries and therefore,
applying the generalized hodograph method,
  we can write the general solution of the Benney
reduction in the implicit form
\begin{equation*}
z^i=v^i x+t,\,\,\,i=1,\dots,n.
\end{equation*}
The inverse scattering solutions found in \cite{giko94,giyu00} are of this form.
 
\section{Hamiltonian structure of the reductions}\label{shamc}
As it was shown in \cite{gits96}, any reduction of Benney is a
diagonalizable and semi-Hamiltonian system of hydrodynamic type.
However, very little is known about the Hamiltonian structure of these systems,
whether local or nonlocal. A few examples are known explicitly. These include the Gelfand-Dikii and Zakharov reductions, which arise as dispersionless limits of known dispersive Hamiltonian systems.
 
In this section, we use Ferapontov's
procedure sketched in Section \ref{hamfor} for semi-Hamiltonian diagonal systems in
order to determine the metric associated with a generic reduction.
\begin{thm}\label{metric}
The general solution of the system  (\ref{meq}) in the case of
Benney reductions
 is
\beq\label{gsmeq} 
g_{ii}=\frac{\de_i A_0}{\varphi_i(\lambda^i)} 
\eeq
 where
 the functions $\varphi_i(\lambda^i)$ are $n$ arbitrary functions of one
variable, the functions $\lm^1,\dots,\lm^n$ being the Riemann invariants of the system.
\end{thm}
 
\noindent \emph{Proof}. From the system (\ref{meq}), and making use
of both the Gibbons-Tsarev equations (\ref{gt}), which hold for any Benney reduction, we obtain
\beq
\de_j\ln{\sqrt{g_{ii}}}=\frac{\de_j
v^i}{v^j-v^i}=\frac{\de_j
A_0}{(v_j-v_i)^2}=\de_j\ln{\sqrt{\de_i A_0}}
\eeq
from which we obtain the general solution (\ref{gsmeq}).
\begin{flushright}
$\Box$
\end{flushright}
 
\vskip 5mm
\noindent 
In the case $\varphi_i=1$ the rotation coefficients
\beq\label{rotcoe}
\beta_{ij}:=\frac{\de_i\sqrt{g_{jj}}}{\sqrt{g_{ii}}}
\eeq
are
symmetric:
\begin{equation}\label{rotcoeex}
\beta_{ij}=\frac{1}{2}\frac{\de_i\de_j A_0}{\sqrt{\de_i A_0\de_j
A_0}}=\frac{\sqrt{\de_i A_0\de_j A_0}}{(v_i-v_j)^2}.
\end{equation}
\noindent
We now focus our attention on this case, that is we consider the
Egorov (potential) metric
\begin{equation}\label{metricH} 
g_{ii}=\de_i A_0.
\end{equation}

\begin{rmk}
We notice that the choice of potential metric is not restrictive, as
any other metric (\ref{gsmeq}) can be written in potential form
under a change of coordinates
\beq\label{chco} \lm^i\longmapsto \varphi_i(\lm^i),
\eeq
which is exactly the freedom we have in the definition of the Riemann invariants.
On the other hand, the choice of the Riemann invariants determines
a unique metric which is potential in those coordinates.
\end{rmk}

In order to find the Christoffel symbols and the curvature tensor of the metric \eqref{metricH}, one could compute these objects -- as usual -- starting from their definitions. However, for a Benney reduction, this procedure can be shortened. Indeed, using the Gibbons-Tsarev system \eqref{gt}, the connection and the curvature can be written as simple algebraic combinations of the quantities
$$v^i, \quad \de_i A^0,\quad \delta(v^i), \quad \delta(\log\sqrt{\de_i A^0}), \qquad i=1,\dots,n,$$ 
where we introduced the shift operator
$$\delta=\sum_{k=1}^n\frac{\de}{\de\lm^k}.$$
We have

\begin{prop}
The symbols
$$\Gamma_k^{ij}=-g^{is}\Gamma_{sk}^j=-\frac{1}{2}g^{is}g^{jl}\Big(\de_s g_{lk}+\de_k g_{ls}-\de_l g_{sk}\Big),$$
where $\Gamma^k_{ij}$ are the Christoffel symbols associated  to the diagonal metric $g_{ii}=\de_i A_0$, for a Benney reduction are given by
\begin{subequations}\label{gamma}
\begin{eqnarray}
&&\Gamma_k^{ij}=0,\hspace{150pt} i\ne j\ne k\\
&&\Gamma_i^{ij}=-\Gamma_i^{ji}=\frac{1}{(v_i-v_j)^2},\hspace{67pt}i\ne j\\
&&\Gamma_k^{ii}=-\frac{\de_k A_0}{\de_i
A_0}\frac{1}{(v_k-v_i)^2},\hspace{70pt}i\ne k \\
&&\notag\\
&&\Gamma_i^{ii}=\sum_{k\neq i}\frac{\de_k A^0}{\de_i A^0}\frac{1}{(v^i-v^k)^2}-\frac{\delta(\ln\sqrt{\de_i A^0})}{\de_i A^0}.
\end{eqnarray}
\end{subequations}
\end{prop}
 
\noindent
\pf For the metric $g_{ij}=\delta_{ij}\de_i A_0$, we get
$$\Gamma^{ij}_k=-\frac{1}{2}\frac{1}{\de_i A^0\de_i A^0}\Big(\delta_{jk}\de_{ik}A^0
+\delta_{ij}\de_{kj}A^0-\delta_{ik}\de_{jk}A^0\Big),$$
and equations \eqref{gamma} are obtained from these by substituting, whenever is allowed, the Gibbons-Tsarev equations \eqref{gt2}.
\begin{flushright}
$\Box$
\end{flushright}

\noindent
In particular, the curvature can be expressed solely via the shift operator $\delta$, acting on $v^i$ and $\ln\sqrt{\de_i A^0}$.
 
\begin{prop}
The non vanishing components of the curvature tensor of the metric
(\ref{metricH}) for a Benney reduction can be
 written in terms of the quantities $\delta(v^i)$, $\delta(\ln{\sqrt{\de_i A^0}})$.
 More precisely, we have the following identity
\begin{equation}\label{curv}
R^{ij}_{ij}=\frac{\delta(\ln{\sqrt{\de_i A^0}})
+\delta(\ln{\sqrt{\de_j A^0}})}{{(v^i-v^j)^2}}-2\,\frac{\delta(v^i)-\delta(v^j)}{(v^i-v^j)^3}.
\end{equation}
\end{prop}
 
\pf Since the rotation coefficients of the metric (\ref{metricH}) are symmetric, it is easy to check that
$$R^{ij}_{ij}=\delta\left(\beta_{ij}\right).$$
Using the Gibbons-Tsarev system \eqref{gt} we get \eqref{curv}. Moreover, it is well known that for a semi-Hamiltonian system the other components of the Riemann tensor are identically zero.
\begin{flushright}
$\Box$
\end{flushright}
 
\vskip 5mm
Formula \eqref{curv} presents a compact way of describing the curvature tensor of the Poisson structure \eqref{NLPB} associated with a Benney reduction. However, we should notice here that the knowledge of the curvature is not sufficient to write the Poisson bracket. Indeed, what we need is a decomposition \eqref{exp}  of the curvature in terms of the symmetries of the system. From formula \eqref{curv} this decomposition looks non-trivial; we will address this problem in the next section using a different approach.

\section{Hamiltonian structure in the $\lm$ picture}\label{shamv}
 
The purpose of this Section is to derive an explicit formulation for the Hamiltonian structure of a reduction of Benney in terms of the function $\lm(p)$, which defines the reduction itself. In particular, we will show how the differential geometric objects associated with Ferapontov's Poisson operator of type \eqref{NLPB} can be expressed, in the case of a Benney reduction with associated $\lm$, in terms of the set of data
$$v^i, \quad \lm^{''}(v^i),\quad \lm^{'''}(v^i),\quad\lm^{''''}(v^i), \qquad\qquad i=1,\dots,n,$$
where $v^i$, the characteristic velocities of the reduction, are the critical points of $\lm$. Moreover, we will describe the quadratic expansion of the curvature associated with the metric.

Let us consider a Benney reduction with associated function $\lm(p)$. In this case, as already mentioned, a set of Riemann invariants is naturally selected, the critical values of $\lm(p)$. From \eqref{2lma} and \eqref{gsmeq}, it is immediate to check that the components of the metric which is potential in those coordinates can be expressed in terms of $\lm$ as
\beq\label{potmeteq}
g_{ii}=\de_i A^0=\frac{1}{\lm^{''}(v^i)}=\underset{p=v^i}{\rm{res}}\left(\frac{dp}{\lm^{'}(p)}\right).
\eeq

This result was already known in the case of dispersionless Gelfand-Dikii \cite{dulizh07} reductions. However,
it holds for all Benney reductions.
 
\subsection{Completing the Loewner system}
We move now our attention from the metric to the Christoffel symbols and the curvature tensor, looking for a way to describing these objects in terms of $\lm$ and its critical points. However, this step is not immediate. Indeed, from equations \eqref{gamma} and \eqref{curv} we need to find an expression in the $\lambda$ picture for the quantities
$$\delta(v^i),\quad \delta(\log\sqrt{\de_i A^0}),$$
we will see that the right object to look at is
\beq\label{funF}
F(p)=\frac{\de p}{\de\lm}+\sum_{i=1}^n\frac{\de p}{\de\lm^i},
\eeq
obtained from the inverse function of $\lm$ with respect to $p$, that is
\beq
p=p(\lm,\lm^1,\dots,\lm^n).
\eeq
The function $F$ is thus determined once the function $\lm(p,\lm^1,\dots,\lm^n)$ is
known. Before discussing the Christoffel symbols and the curvature, we will consider the properties of this function in detail. First of all, using the Loewner equations (\ref{loewp}) and the expression \eqref{2lma}, we can write $F(p)$ in the following form:
\begin{align}
F(p)&=\frac{1}{\lm^{'}(p)}-\sum_{i=1}^n\frac{\de_i A^0}{p-v^i}\notag\\
&=\frac{1}{\lm^{'}(p)}-\sum_{i=1}^n
\underset{p=v_i}{\rm{res}}\left(\frac{1}{\lm^{'}(p)}\right)\frac{1}{p-v^i}.\label{analiticlm}
\end{align}
From its expansion \eqref{oneoverlm}, the function $\frac{1}{\lm^{'}(p)}$
in $v^i$ has a simple pole, therefore, $F(p)$ is analytic at $p=v^i$. Using this fact, we can prove the following
 
\begin{thm}\label{metriclambda}
Let $\lm(p,\lm^1,\dots,\lm^n)$ be a solution of equation (\ref{vlasov}), and let $v^1,\dots,v^n$ be its critical points.
Defining the function $F(p)$ as above (\ref{funF}), we have
\begin{align}
F(v^i)&=\delta(v^i)\label{fdelv}\\
&\notag\\
\frac{\de F}{\de p}(v^i)&=\delta(\ln{\sqrt{\de_i A^0}}).\label{fdela}
\end{align}
\end{thm}
 
\noindent
\pf We have already shown that $F(p)$ is analytic at $p=v^i$.
In order to prove (\ref{fdelv}) and (\ref{fdela}), we consider the system of $n+1$ differential equations
$$$$
\begin{align}\label{genkk}
\frac{\de p}{\de\lm^i}&=\frac{\de_i A^0}{v^i-p}\qquad\qquad i=1,\dots,n\notag\\
\\
\frac{\de p}{\de\lm}&=\sum_{k=0}^n\frac{\de_k A^0}{p-v^k}+F(p).\notag
\end{align}
The conditions
\beq
\frac{\de^2 p}{\de\lm^i\de\lm^j}-\frac{\de^2 p}{\de\lm^j\de\lm^i}=0 \qquad
i\neq j
\eeq
give nothing but the Gibbons-Tsarev system, hence are satisfied for any reduction.
So, we concentrate our attention on the remaining $n$ consistency conditions
\beq\label{cc}
\frac{\de^2 p}{\de\lm\de\lm^i}-\frac{\de^2 p}{\de\lm^i\de\lm}=0,
\eeq
which -- by construction -- are satisfied, to obtain some information about $F(p)$. Expanding both sides we obtain
\begin{align*}
\frac{\de^2 p}{\de\lm\de\lm^i}=&\frac{\de_i A_0}{(p-v_i)^2}\frac{\de p }{\de\lm}\\
=&\frac{\de_i A_0}{(p-v_i)^2}\left[F(p)
+\sum_{k=1}^n\frac{\de_k A_0}{p-v_k}\right],
&\\
&\\
\frac{\de^2 p}{\de\lm^i\de\lm}=&\frac{\de F}{\de p}\,\frac{\de p}{\de\lm^i}+\frac{\de F}{\de \lm^i}+
\sum_{k=1}^n\left[\frac{\de_k\de_i A_0}{p-v_k}+\frac{\de_k A_0}{(p-v_k)^2}\left(\de_i v_k-\frac{\de p}{\de\lm^i}\right)\right]\\
=&\frac{\de F}{\de p}\,\frac{\de_iA^0}{v^i-p}+\frac{\de F}{\de \lm^i}+
\sum_{k=1}^n\left[\frac{\de_k\de_i A_0}{p-v_k}+\frac{\de_k A_0}{(p-v_k)^2}\left(\de_i v_k+\frac{\de_iA^0}{p-v^i}
\right)\right],
\end{align*}
substituting the Gibbons-Tsarev equations (\ref{gt}) in the above formulae and rearranging
(\ref{cc}), we find that
 
\beq\label{ccf}
\frac{F(p)-\delta(v^i)}{(p-v^i)^2}+\frac{\frac{\de F}{\de p}(p)-2\delta(\log{\sqrt{\de_i
A^0}})}{p-v^i}-\frac{1}{\de_i A^0}\frac{\de F(p)}{\de\lm^i}=0.
\eeq
 
Multiplying by $(p-v^i)^2$ and taking the limit for $p\to v^i$, we get
\beq
\label{uu}
F(v^i)=\delta(v^i)\notag.
\eeq
Then, taking the residue of the right hand side of \eqref{ccf} at $p=v^i$ gives
\beq
\frac{\de F}{\de p}(v^i)=\delta(\log{\sqrt{\de_i A^0}}).\notag
\eeq
\begin{flushright}
$\Box$
\end{flushright}
\noindent

Thus, specifying the function $F$ turns out to be the analogue, in the $\lm$ picture, of completing the system (\ref{gt}), from which we were able to
express the Christoffel symbols and the curvature tensor of the metric.

It follows from its definition, that $F(p)$ is obtained from $\frac{1}{\lm^{'}(p)}$ by subtracting off its singularities  at $p=v^i$. The importance of this function is that it describes the invariant properties of the reduction. For instance, if a reduction admits a function $\lm$ associated with it such that $F(p)=1$, then the reduction is Galilean invariant. The case $F(p)=p$ corresponds instead to the scaling invariance of the system. As seen in example below, the function $F$, hence these invariances, are strongly related with the curvature of the Poisson bracket.
 
\begin{exam}
The $2-$component Zakharov reduction is known to possess both Galilean and scaling invariance. Using the technique above, the former can be explained by saying that, for the function $\lm(p)$ given in Example \ref{lmzak}, it follows $F(p)=1$. Thus, we get 
$$\delta(v^i)=\sum_{k=1}^n\frac{\de v^i}{\de \lm^k}=1,$$ 
where the Riemann invariants are the critical values of $\lm(p)$. For the scaling invariance, one can proceed as follows: define the function $\varphi(p)=\ln{\lm(p)}$,  where $\lm(p)$ is the same as before. This new function has the same critical points $v^i$ as $\lm$, plus the poles of $\lm(p)$, which play no role in the deerivation of the reduction (see Remark \ref{extraz}). Hence, it is associated with the Zakharov reduction, and in this case $F(p)=p$. Thus
\beq\label{dex}
\delta(v^i)=\sum_{k=1}^n\frac{\de v^i}{\de \varphi^k}=v^i,
\eeq
where the $\varphi^i=\ln{\lm^i}$ are the natural Riemann invariants associated with $\varphi(p)$, i.e. its critical values. It is elementary to show that \eqref{dex} corresponds to the scaling invariance with respect to the $\lm^i$.
\end{exam}

\begin{rmk}
System (\ref{genkk}) was considered for the first time in connection with a Benney reduction by Kokotov
and Korotkin \cite{koko07}, in the particular case of the $N-$component Zakharov reduction, where $F(p)\!=1$. 
\end{rmk}
 
In the next section, we will use the function $F$ to describe the Christoffel symbols and the curvature in terms of $\lm$. Nevertheless, the latter can be expressed directly in terms of $F$ using the following residue formula

\begin{prop}\label{curvf}
In terms of the function $F$, the non vanishing components
of the Riemann
tensor satisfy the following identity:
\begin{equation}
R^{ij}_{ij}= \sum_{k=i,j}\underset{p=v_k}{\rm
res}\left(\frac{F(p)}{(p-v^i)^2(p-v^j)^2}\,dp\right)=\f{1}{2\pi
i}\int_{\Gamma_i\cup\,\Gamma_j}\frac{F(p)}{(p-v^i)^2(p-v^j)^2}\,dp
\end{equation}
where $\Gamma_i$ and $\Gamma_j$ are two sufficiently small
contours around $p=v^i$ and $p=v^j$.
\end{prop}
\vskip 5mm
\noindent
\emph{Proof}. \label{mainid2} From (\ref{curv}) and
Theorem \ref{metriclambda} it follows
immediately that
\beq
R^{ij}_{ij}=\frac{1}{(v^i-v^j)^2}\left[\left(\frac{\de
F}{\de p}(v^i)+ \frac{\de
F}{\de
p}(v^j)\right)-2\,\frac{F(v^i)-F(v^j)}{v^i-v^j}\right]
\eeq
It is easy to check the chain of identities:
\begin{gather*}
\frac{1}{(v^i-v^j)^2}\left[\left(\frac{\de F}{\de p}(v^i)+
\frac{\de
F}{\de
p}(v^j)\right)-2\,\frac{F(v^i)-F(v^j)}{v^i-v^j}\right]=\\
\lim_{p\to
v^i}\frac{d}{dp}\left[\frac{F(p)}{(p-v^j)^2}\right]+\lim_{p\to
v^j}\frac{d}{dp}\left[\frac{F(p)}{(p-v^i)^2}\right]=\\
\lim_{p\to
v^i}\frac{d}{dp}\left[(p-v^i)^2\frac{F(p)}{(p-v^j)^2(p-v^i)^2}\right]
+\lim_{p\to
v^j}\frac{d}{dp}\left[(p-v^j)^2\frac{F(p)}{(p-v^i)^2(p-v^j)^2}\right]=\\
\underset{p=v^i}{\rm
res}\left(\frac{F(p)}{(p-v^i)^2(p-v^j)^2}\,dp\right)+\underset{p=v^j}{\rm
res}\left(\frac{F(p)}{(p-v^i)^2(p-v^j)^2}\,dp\right).
\end{gather*}
In the last identity we used the fact that $F(p)$ is
regular
at $p=v^i$, for all $i$.
\begin{flushright}
$\Box$
\end{flushright}

\subsection{Christoffel symbols and Curvature tensor}
\subsubsection{Potential metric}
We are now able to complete the description of the Poisson bracket associated with a Benney reduction, in the case where the metric is potential in the coordinate used.
 
\begin{prop}
The Christoffel symbols \eqref{gamma} of the potential metric \eqref{potmeteq} can be written, in terms of the function $\lm$, as
\begin{subequations}\label{gammalm}
\begin{eqnarray}
&&\Gamma_k^{ij}=0,\hspace{155pt} i\ne j\ne k\\
&&\Gamma_i^{ij}=-\Gamma_i^{ji}=\frac{1}{(v^i-v^j)^2},\hspace{72pt}i\ne j\\
&&\Gamma_k^{ii}=-\frac{\lm^{''}(v^i)}{\lm^{''}(v^k)}\frac{1}{(v_k-v^i)^2},\hspace{70pt}i\ne k \\
&&\notag\\
&&\Gamma_i^{ii}=\frac{1}{6}\frac{\lm^{''''}(v^i)}{\lm^{''}(v^i)}-\frac{1}{4}\frac{\lm^{'''}(v^i)^2}{\lm^{''}(v^i)^2},\label{schchr}
\end{eqnarray}
\end{subequations}
The curvature tensor \eqref{curv} of the potential metric \eqref{potmeteq} can be written, in terms of the function $\lambda(p)$ as
\begin{align}\label{curvlm}
R^{ij}_{ij}=&\frac{3}{(v^i-v^j)^4}\left(\frac{1}{\lm^{''}(v^i)}+\frac{1}{\lm^{''}(v^j)}\right)+
\frac{1}{(v^i-v^j)^3}\left(\frac{\lm^{'''}(v^i)}{\lm^{''}(v^i)^2}-\frac{\lm^{'''}(v^j)}{\lm^{''}(v^j)^2}\right)\notag\\
&+\frac{1}{(v^i-v^j)^2}\left(\frac{1}{4}\frac{\lm^{'''}(v^i)^2}{\lm^{''}(v^i)^3}-\frac{1}{6}\frac{\lm^{''''}(v^i)}{\lm^{''}(v^i)^2}+
\frac{1}{4}\frac{\lm^{'''}(v^j)^2}{\lm^{''}(v^j)^3}-\frac{1}{6}\frac{\lm^{''''}(v^j)}{\lm^{''}(v^j)^2}\right)\notag\\
&+\sum_{k\neq
i,j}\frac{1}{\lm^{''}(v^k)}\frac{1}{(v^i-v^k)^2(v^j-v^k)^2}.
\end{align}
\end{prop}
 
\pf Starting from the definition \eqref{funF} of $F$, and using \eqref{oneoverlm}, we can write
\begin{align}
F(v^i)&=-\frac{1}{2}\frac{\lm^{'''}(v^i)}{\lm^{''}(v^i)^2}-\sum_{k\neq
i}\frac{1}{\lm^{''}(v^k)}\frac{1}{v^i-v^k},\\
&\notag\\
\frac{\de F}{\de p}(v^i)&=\frac{1}{4}\frac{\lm^{'''}(v^i)^2}{\lm^{''}(v^i)^3}
-\frac{1}{6}\frac{\lm^{''''}(v^i)}{\lm^{''}(v^i)^2}+
\sum_{k\neq
i}\frac{1}{\lm^{''}(v^k)}\frac{1}{(v^i-v^k)^2}.
\end{align}
Then, by Theorem \ref{metriclambda}, and substituting the above expressions for $F$ into \eqref{gamma} and \eqref{curv}, we obtain \eqref{gammalm} and \eqref{curvlm} respectively.
\begin{flushright}
$\Box$
\end{flushright}

\begin{rmk}
We note that \eqref{schchr} is a constant multiple of the Schwarzian derivative of $\lm^{'}(p)$, evaluated at $p=v^i$. 
\end{rmk}

Recalling the expression \eqref{potmeteq}  for the the metric, form the Proposition above we find that the whole Poisson operator associated with a Benney reduction of symbol $\lm$ depends only on the critical points of $\lm$ and on the value of its second, third and fourth derivatives evaluated at these points.
 
We have now all we need to write the nonlocal
tail of the Hamiltonian structure associated
  to the metric $g_{ii}=\d_i A_0$.
  
\begin{prop}
The non-vanishing components of the Riemann tensor of the metric \eqref{potmeteq} admit
  the following quadratic expansion
\begin{equation}\label{mainid3}
R^{ij}_{ij}=\frac{1}{2\pi i}\int_{C} w^i(\lm) w^j(\lm)
d\lm.
\end{equation}
where $C=C_1\cup\dots\cup C_n$ with $C_i$ described as above, and  the functions
$$w^i(\lambda)=\frac{\frac{\de
p}{\de\lambda}}{(p(\lambda)-v^i)^2},$$
are the generating functions of the symmetries \eqref{symgenfun}.
Consequently the non local tail of the Hamiltonian
structure associated
  to the metric $g_{ii}=\d_i A_0$ is given by
$$\frac{1}{2\pi i}\int_C
w^i(\lm)\lm^i_x\,\,\left(\frac{d}{dx}\right)^{-1}\!\!w^j(\lm)\lm^j_xd\lm.$$
\end{prop}
 
\vskip 5mm
\pf We prove the Proposition showing that the integral in (\ref{mainid3}) is the same as the right hand side of \eqref{curvlm}. First,
writing the integral
$$\frac{1}{2\pi i}\int_{C} w^i(\lm) w^j(\lm) d\lm$$ in
terms of the variable $p$ we obtain
\begin{equation}
\label{mainid4}
R^{ij}_{ij}=\frac{1}{2\pi i}\int_\Gamma
\frac{\frac{1}{\lm^{'}(p)}}{(p-v^i)^2(p-v^j)^2} dp =
\sum_{k=1}^n\underset{p=v_k}{\rm
res}\left(\frac{\frac{1}{\lm^{'}(p)}}
{(p-v^i)^2(p-v^j)^2}\,dp\right).
\end{equation}
Using \eqref{oneoverlm}, the integrand can be expanded, for $k=1.\dots,n$, as
 
$$
\!\frac{1}{(p-v^i)^2(p-v^j)^2}\left(\frac{1}{\lm^{''}(v^k)}\frac{1}{p-v^k}-
\frac{1}{2}\frac{\lm^{'''}(v^k)}{\lm^{''}(v^k)^2}\!+\!
\left(\frac{1}{4}\frac{\lm^{'''}(v^k)^2}{\lm^{''}(v^k)^3}-\frac{1}{6}\frac{\lm^{''''}(v^k)}{\lm^{''}(v^k)^2}\right)
\!(p-v^k)\!+\!\dots\!\right).
$$
 
Thus, for $k\neq i ,j$ we get
$$\underset{p=v_k}{\rm res}\left(\frac{\frac{1}{\lm^{'}(p)}}
{(p-v^i)^2(p-v^j)^2}\,dp\right)=\frac{1}{\lm^{''}(v^k)}\frac{1}{(v^k-v^i)^2(v^k-v^j)^2},$$
while
\begin{align*}
\underset{p=v^i}{\rm res}\left(\frac{\frac{1}{\lm^{'}(p)}}
{(p-v^i)^2(p-v^j)^2}\,dp\right)=&\frac{3}{(v^i-v^j)^4}\frac{1}{\lm^{''}(v^i)}+
\frac{1}{(v^i-v^j)^3}\frac{\lm^{'''}(v^i)}{\lm^{''}(v^i)^2}\\
&+\frac{1}{(v^i-v^j)^2}\left(\frac{1}{4}\frac{\lm^{'''}(v^i)^2}{\lm^{''}(v^i)^3}
-\frac{1}{6}\frac{\lm^{''''}(v^i)}{\lm^{''}(v^i)^2}\right),\\
&\\
\underset{p=v^j}{\rm res}\left(\frac{\frac{1}{\lm^{'}(p)}}
{(p-v^i)^2(p-v^j)^2}\,dp\right)=&\frac{3}{(v^j-v^i)^4}\frac{1}{\lm^{''}(v^j)}+
\frac{1}{(v^j-v^i)^3}\frac{\lm^{'''}(v^j)}{\lm^{''}(v^j)^2}\\
&+\frac{1}{(v^j-v^i)^2}\left(\frac{1}{4}\frac{\lm^{'''}(v^j)^2}{\lm^{''}(v^j)^3}
-\frac{1}{6}\frac{\lm^{''''}(v^j)}{\lm^{''}(v^j)^2}\right),
\end{align*}
From these, formula \eqref{curvlm} for the curvature tensor follows. The last statement of the Proposition is a consequence of
the general theory of Ferapontov.
\begin{flushright}
$\Box$
\end{flushright}
 
\begin{rmk}
Alternatively, one can prove the above result using starting from the function $F$, namely deforming the integral
$$\f{1}{2\pi i}\int_{\Gamma_i\cup\,\Gamma_j}\frac{F(p)}{(p-v^i)^2(p-v^j)^2}\,dp$$
which is shown in Proposition \ref{curvf} to be equal to the curvature,   into \eqref{mainid3}. In order to do so, it is sufficient to verify the following identity
\begin{eqnarray*}
&&-\frac{1}{2\pi
i}\int_{\Gamma_i\cup\Gamma_j}\frac{\sum_{k=1}^n\frac{\partial_k
A_0}{p-v^k}}{(p-v^i)^2(p-v^j)^2}dp=\frac{1}{2\pi
i}\int_{\Gamma-(\Gamma_i\cup\Gamma_j)}\frac{\frac{1}{\frac{\partial\lambda}{\partial
p}}}{(p-v^i)^2(p-v^j)^2}dp,
\end{eqnarray*}
that can be proved by straightforward computation. Indeed,
since the left hand side is the sum of residues at $p=v^i$
and $p=v^j$
  of a function having a pole of order 3 at $p=v^i,v^j$ we
obtain
\begin{eqnarray*}
&&l.h.s=-\left(\lim_{p\to v^i}+\lim_{p\to
v^j}\right)\frac{1}{2}\frac{d^2}{d
p^2}\sum_{k=1}^n\left[\frac{(p-v^i)\partial_k
A_0}{(p-v^j)^2(p-v^k)}\right]\\
&&=\sum_{k\ne i,j}\f{\d_k
A_0}{(v^i-v^j)^2}\left[\frac{2}{(v^i-v^j)(v^i-v^k)}
+\frac{1}{(v^i-v^k)^2}-\frac{2}{(v^i-v^j)(v^j-v^k)}+\frac{1}{(v^j-v^k)^2}\right]\\
&&=\sum_{k\ne i,j}\frac{\partial_k
A_0}{(v^i-v^k)^2(v^j-v^k)^2}.
\end{eqnarray*}
On the other hand the right hand side is the sum of
residues at $p=v^k$, for $k\ne i,j$
  of a function having simple poles at $p=v^k$:
\begin{eqnarray*}
r.h.s.=\sum_{k\ne i,j}\lim_{p\to
v^k}\frac{(p-v^k)\frac{1}{\frac{\partial\lambda}{\partial
p}}}{(p-v^i)^2(p-v^j)^2}=\sum_{k\ne i,j}\frac{\partial_k
A_0}{(v^i-v^k)^2(v^j-v^k)^2}=l.h.s.
\end{eqnarray*}
\end{rmk}
 
\vskip 10mm
\noindent 
Recalling the results above, we have the following
\begin{thm}
The reduction of Benney associated with the function $\lm(p,\lm^1,\dots,\lm^n)$
is Hamiltonian with the Hamiltonian structure 
\beq\label{potpoibra}
\Pi^{ij}=\lm^{''}(v^i)\delta^{ij}\frac{d}{dx}+\Gamma^{ij}_k\,\lm^k_x+\frac{1}{2\pi i}\int_{C}\frac{\frac{\de
p}{\de\lm}\,\,\,\lm^i_x}{(p(\lm)-v^i)^2}\left(\frac{d}{d
x}\right)^{-1}\!\!\!\frac{\frac{\de
p}{\de\lm}\,\,\,\lm^j_x}{(p(\lm)-v^j)^2}\,d\lm,
\eeq
where
\begin{align*}
\Gamma^{ij}_k\,\lm^k_x&=\frac{\lm^i_x-\lm^j_x}{(v^i-v^j)^2}\qquad\qquad i\neq j,\\
&\\
\Gamma^{ii}_k\,\lm^k_x&=\left(\frac{1}{6}\frac{\lm^{''''}(v^i)}{\lm^{''}(v^i)}
-\frac{1}{4}\frac{\lm^{'''}(v^i)^2}{\lm^{''}(v^i)^2}
\right)\lm^i_x-\sum_{k\neq i}\frac{\lm^{''}(v^i)}{\lm^{''}(v^k)}
\frac{\lm^k_x}{(v^i-v^k)^2}
\end{align*}
and $C=C_1\cup\dots\cup C_n$. Here, the $v^i$ are the critical points of $\lm$, and the $\lm^i$ the critical values. In this coordinates, the metric $g_{ij}=\frac{\delta^{ij}}{\lm^{''}(v^i)}$ is potential.
\end{thm}

\vskip 15mm
\subsubsection{The general case}
As we pointed out previously, any metric \eqref{gsmeq} associated with a reduction can be put in potential form, after a suitable change of the Riemann invariants. However, it is often convenient to write the expression of the Poisson operators generated by these metrics in terms of the Riemann invariants selected by $\lambda$. Thus, we consider the metrics
\beq\label{metricfi}
g_{ii}=\frac{\de_i
A^0}{\varphi_i(\lm^i)}=\frac{1}{\varphi_i(\lm^i)\lm^{''}(v^i)},
\eeq
where $\varphi_i\!=\!\varphi_i(\lm^i)$ are
arbitrary functions, and we proceed as before.
 
\begin{prop}
The Christoffel symbols appearing in the Hamiltonian
structure are given by
 
\begin{eqnarray*}
&&\Gamma_k^{ij}=\,0,\hspace{150pt} i\ne j\ne k\\
&&\Gamma_i^{ij}=\,\frac{\varphi_j}{(v^i-v^j)^2},\hspace{106pt}i\ne
j\\
&&\Gamma_j^{ij}=\,-\frac{\varphi_i}{(v^i-v^j)^2},\hspace{97pt}i\ne
j\\
&&\Gamma_k^{ii}=\,-\frac{\lm^{''}(v^i)}{\lm^{''}(v^k)}\frac{\varphi_i}{(v^k-v^i)^2},\hspace{63pt}i\ne k \\
&&\Gamma_i^{ii}=\,\varphi_i\,\left(\frac{1}{6}\frac{\lm^{''''}(v^i)}{\lm^{''}(v^i)}
-\frac{1}{4}\frac{\lm^{'''}(v^i)^2}{\lm^{''}(v^i)^2}\right)+\frac{1}{2}\varphi^{'}_i.
\end{eqnarray*}
Here we denote
$$\varphi^{'}_i=\frac{d\varphi}{d\lm^i}(\lm^i).$$
The nonlocal tail appearing in the Hamiltonian structure
is then given by
\begin{equation}
\label{intfi}
\frac{1}{2\pi i}\sum_{k=1}^n\int_{C_k}
\frac{\frac{\de p}{\de\lm}\,\,\,\lm^i_x
}{(p(\lm)-v^i)^2}\left(\frac{d}{dx}\right)^{-1}\!\!\frac{\frac{\de
p}{\de\lm}\,\,\,\lm^j_x}{(p(\lm)-v^j)^2}\,\,\varphi_k(\lm)\, d\lm.
\end{equation}
\end{prop}
 
\vskip 5mm
\pf The proof of the formula for the $\Gamma^{ij}_k$ is a straightforward
computation. Let us prove the second statement. Since nothing new is
involved in such computations
we will skip the details. For the metric \eqref{metricfi}, the non vanishing components of the curvature tensor are
\begin{align}\label{curvfi}
R^{ij}_{ij}=&\,\,\frac{3}{(v^i-v^j)^4}\left(\varphi_i\,\de_i
A^0+\varphi_j\,\de_j A^0\right)
-\,\frac{2}{(v^i-v^j)^3}\left(\varphi_i\,\de_i
v^i-\varphi_j\,\de_j v^j\right)\notag\\
&+\frac{1}{(v^i-v^j)^2}\left(\varphi_i\,\de_i\ln\sqrt{\de_i
A^0}+\varphi_j\,\de_j\ln\sqrt{\de_j A^0}\right)\notag\\
&+\sum_{k\neq i,j}\frac{\varphi_k\,\de_k
A^0}{(v^i-v^k)^2(v^j-v^k)^2}
+\frac{1}{2}\,\frac{\varphi^{'}_i+\varphi^{'}_j}{(v^i-v^j)^2}.
\end{align}
Expression \eqref{curvfi} can be written, in terms of
$\lm$, as
\begin{align}\label{curvlmfi}
R^{ij}_{ij}=&\frac{3}{(v^i-v^j)^4}\left(\frac{\varphi_i}{\lm^{''}(v^i)}+\frac{\varphi_j}{\lm^{''}(v^j)}\right)+
\frac{1}{(v^i-v^j)^3}\left(\varphi_i\,\frac{\lm^{'''}(v^i)}{\lm^{''}(v^i)^2}
-\varphi_i\,\frac{\lm^{'''}(v^j)}{\lm^{''}(v^j)^2}\right)\notag\\
&+\frac{1}{(v^i-v^j)^2}\left(\varphi_i\left(\frac{1}{4}\frac{\lm^{'''}(v^i)^2}{\lm^{''}(v^i)^3}
-\frac{1}{6}\frac{\lm^{''''}(v^i)}{\lm^{''}(v^i)^2}\right)+
\varphi_j\left(\frac{1}{4}\frac{\lm^{'''}(v^j)^2}{\lm^{''}(v^j)^3}
-\frac{1}{6}\frac{\lm^{''''}(v^j)}{\lm^{''}(v^j)^2}\right)\right)\notag\\
&+\sum_{k\neq
i,j}\frac{1}{\lm^{''}(v^k)}\frac{\varphi_k}{(v^i-v^k)^2(v^j-v^k)^2}+
\frac{1}{2}\,\frac{\varphi^{'}_i+\varphi^{'}_j}{(v^i-v^j)^2},
\end{align}
The equivalence between \eqref{intfi} and the right hand
side of \eqref{curvlmfi} can be obtained by rewriting the
integrals above in the $p-$plane,
\beq
\frac{1}{2\pi
i}\sum_{k=1}^n\int_{\Gamma_k}\frac{\varphi_k(\lm(p))\,\,\frac{1}{\lm^{'}(p)}}{(p-v^i)^2(p-v^j)^2}
dp.
\eeq
and using the same arguments of the main theorem, except
that for every $k$, in the integral around $\Gamma_k$ we
have to consider also the contribution of the function
$\varphi_k(\lm(p))$ which expands, at $p=v^k$, as
$$\varphi_k(\lm(p))=\varphi_k\,+\,\frac{\varphi^{'}_k}{2}\,\lm^{''}(v^k)\,(p-v^k)^2+\dots$$
\begin{flushright}
$\Box$
\end{flushright}

\noindent
It follows from the above that we have
\begin{thm} 
The reduction of Benney associated with the function $\lm(p,\lm^1,\dots,\lm^n)$
is Hamiltonian with the family of Hamiltonian structures 
\begin{align}\label{npotpoibra}
\Pi^{ij}=&\,\,\varphi_i\,\lm^{''}(v^i)\delta^{ij}\frac{d}{dx}+\Gamma^{ij}_k\,\lm^k_x\notag\\
&\notag\\
&+\frac{1}{2\pi i}\sum_{k=1}^n\int_{C_k}\frac{\frac{\de
p}{\de\lm}\,\,\,\lm^i_x}{(p(\lm)-v^i)^2}\left(\frac{d}{d
x}\right)^{-1}\!\!\!\frac{\frac{\de
p}{\de\lm}\,\,\,\lm^j_x}{(p(\lm)-v^j)^2}\,\varphi_k(\lm)\,d\lm,
\end{align}
with
\begin{align*}
\Gamma^{ij}_k\,\lm^k_x&=\frac{\varphi_j\,\lm^i_x-\varphi_i\,\lm^j_x}{(v^i-v^j)^2}\qquad\qquad i\neq j,\\
&\\
\Gamma^{ii}_k\,\lm^k_x&=\varphi_i\,\left(\frac{1}{6}\frac{\lm^{''''}(v^i)}{\lm^{''}(v^i)}
-\frac{1}{4}\frac{\lm^{'''}(v^i)^2}{\lm^{''}(v^i)^2}
\right)\lm^i_x+\frac{1}{2}\varphi^{'}_i\,\lm^i_x-\sum_{k\neq i}\frac{\lm^{''}(v^i)}{\lm^{''}(v^k)}
\frac{\varphi_i\,\lm^k_x}{(v^i-v^k)^2},
\end{align*}
where $\varphi_1,\dots,\varphi_n$ are arbitrary functions of a single variable. Here, the $v^i$ are the critical points of $\lm$, the coordinates $\lm^i$ the corresponding critical values, and the $C_k$ are the contours defined above.
\end{thm}

\section{Finite nonlocal tail: some examples}
In the expression \eqref{npotpoibra} for the Poisson operator, the components of the curvature are expressed as integrals of functions around suitable contours in a complex domain. A natural question to ask is whether this integral can be reduced to a finite sum, and we will show now some examples where this is possible. For simplicity, we will consider the case when 
$\varphi_1(\lm)=\dots=\varphi_n(\lm)=\lm^k$, for $k\in\mathbb{Z}$. In this case the curvature can be expressed as
\begin{equation*}
R^{ij}_{ij}=\frac{1}{2\pi i}\,\int_C\frac{\left(\frac{\de p}{\de\lm}\right)^2\,\lm^k}
{(p(\lm)-v^i)^2(p(\lm)-v^j)^2}\,d\lm, \qquad k\in\mathbb{Z}.
\end{equation*}
Essentially, the finite expansion appears whenever is possible to substitute the the contour $C$ with a contour a round $\lm=\infty$ and a finite number of other marked points. We illustrate this special situation in two simple examples.

\subsection{2-component Zakharov reduction}

In this case (see examples \ref{exzak1}, \ref{lmzak}), since $\lambda$ is a single-valued rational function of $p,$ it is convenient to work in the $p$-plane. In order to calculate the curvature, the non vanishing components of the Riemann tensor are given by
\begin{equation*}
R^{12}_{12}=\sum_{i=1}^2\underset{p=v^i}{\rm{res}}\left(\frac{\lambda(p)^k\,\,\,\frac{1}{\lm^{'}(p)}}{(p-v_1)^2(p-v_2)^2}\,dp\right), \qquad k\in\mathbb{Z}.
\end{equation*}
The abelian differential
\begin{equation*}
\frac{\lambda(p)^k\,\,\,\frac{1}{\lm^{'}(p)}}{(p-v_1)^2(p-v_2)^2}\,dp
\end{equation*}
has poles at the points  $p=v_1,\,\,$ $p=v_2,\,\,$ as well as:\\
\\
if $k>2$  
$$p=\infty,\quad p=\frac{A_1}{A_0}\hspace{8.5cm}(\text{poles of }
\lambda)$$
if $k<0$
$$\,\,p=s_1=\f{1}{2}\f{A_1+(A_1^2-4A_0^3)^{1/2}}{A_0},\,\,p=s_2=\f{1}{2}\f{A_1-(A_1^2-4A_0^3)^{1/2}}{A_0} \hspace{0.5cm}(\text{zeros of } \lambda),$$

\noindent
while for $k=0,1,2$ there are no other poles. Since the sum of the residues of an abelian differential on a compact Riemann surface is zero, we can substitute the sum of residues at $p=v_1,v_2\,\,$ with, respectively\\

- zero $\quad$ if $\,\,k=0,1,2,$

- minus the sum of residues at
$p=\infty,\,\,$ $p=\frac{A_1}{A_0}\quad$ if $\,\,k>2$

- minus the sum of residues at
$p=s_1,\,\,$ $p=s_2,\quad$ if $\,\,k<0$.\\

\noindent
Summarizing, we have

\begin{align*}
&R^{ij}_{ij}=0, &k=0,1,2,\\
&R^{ij}_{ij}=-\left(\underset{p=\infty}{\rm{res}}+\underset{p=\frac{A^1}{A^0}}{\rm{res}}\right)
\frac{\lambda(p)^k\,\frac{1}{\lm^{'}(p)}}{(p-v^i)^2(p-v^j)^2}\,dp,&k>2,\\
&R^{ij}_{ij}=-\left(\underset{p=s_1}{\rm{res}}+\underset{p=s_2}{\rm{res}}\right)
\frac{\lambda(p)^k\,\frac{1}{\lm^{'}(p)}}{(p-v^i)^2(p-v^j)^2}\,dp,&k<0.
\end{align*}

\noindent
Moreover, as a counterpart in the
$\lambda$-plane of the above formulae we have
\begin{align*}
&R^{ij}_{ij}=0, &k=0,1,2,\\
&R^{ij}_{ij}=-2\,\underset{\lm=\infty}{\rm{res}}\Big(w^1(\lambda)w^2(\lambda)\lambda^k\,d\lambda\Big),&k>2,\\
&R^{ij}_{ij}=-2\,\underset{\lm=0}{\rm{res}}\,\Big(w^1(\lambda)w^2(\lambda)\lambda^k\,d\lambda\Big),&k<0,
\end{align*}
and this shows that the residues have to be computed around marked points, which not depend on the dynamics of the reduction. Expanding $w^1(\lambda)$ and $w^2(\lambda)$ near $\lambda=\infty,$ we get
\begin{eqnarray*}
&&w^1(\lambda)=-\sum_{k=1}^{\infty}\frac{k\,w^1_{k}}{\lambda^{k+1}}=\\
&&-\f{1}{\lambda^2}-\f{2\,v^1}{\lambda^3}-\f{3(2A_0^3+A_1^2+2A_1
A_0^{\f{3}{2}})}{A_0^2}\f{1}{\lambda^4}-
\f{4(A_1^3+6A_1 A_0^3+3A_0^{\f{9}{2}}+3A_1^2
A_0^\f{3}{2})}{A_0^3}\f{1}{\lambda^5}+\dots\\
&&w^2(\lambda)=-\sum_{k=1}^{\infty}\f{k\,w^2_{k}}{\lambda^{k+1}}=\\
&&-\f{1}{\lambda^2}-\f{2\,v^2}{\lambda^3}-\f{3\,(2A_0^3+A_1^2-2A_1
A_0^{\f{3}{2}})}{A_0^2}\f{1}{\lambda^4}
-\f{4\,(A_1^3+6A_1 A_0^3-3A_0^{\f{9}{2}}-3A_1^2
A_0^\f{3}{2})}{A_0^3}\f{1}{\lambda^5}+\dots
\end{eqnarray*}
and near $\lambda=0$
\begin{eqnarray*}
w^1(\lambda)&=&\sum_{h=0}^{\infty}z^1_{-h}\,\lambda^h=
-\f{2A_0^2(-\sqrt{A_1^2-4A_0^3}+A_1)}{\left(A_1-\sqrt{A_1^2-4A_0^3}+2A_0^{\f{3}{2}}\right)^2\sqrt{A_1^2-4A_0^3}}+\\
&&-\f{8A_0^3\left(\sqrt{A_1^2-4A_0^3}(A_0^3-A_1^2)+A_1^3-3A_1
A_0^3+2A_0^{\f{9}{2}}
\right)}{(A_1^2-4A_0^3)^{\f{3}{2}}
\left(A_1-\sqrt{A_1^2-4A_0^3}+2A_0^{\f{3}{2}}\right)^3
}\lambda+\dots\\
w^2(\lambda)&=&\sum_{h=0}^{\infty}z^2_{-h}\,\lambda^h=
-\f{2A_0^2(-\sqrt{A_1^2-4A_0^3}+A_1)}{\left(A_1-\sqrt{A_1^2-4A_0^3}-2A_0^{\f{3}{2}}\right)^2\sqrt{A_1^2-4A_0^3}}+\\
&&-\f{8A_0^3\left(\sqrt{A_1^2-4A_0^3}(A_0^3-A_1^2)+A_1^3-3A_1
A_0^3-2A_0^{\f{9}{2}}
\right)}{(A_1^2-4A_0^3)^{\f{3}{2}}
\left(A_1-\sqrt{A_1^2-4A_0^3}-2A_0^{\f{3}{2}}\right)^3
}\lambda+\dots
\end{eqnarray*}
and taking into account that the coefficients of the
expansion are characteristic velocities of symmetries,
we easily obtain the quadratic expansion of the Riemann
tensor. For $k>2$ we have
$$R^{12}_{12}=\sum_{i+j=k-1}\left(w^1_i w^2_j+w^1_j
w^2_i\right),$$
while for $k<0,\,\,$ we obtain
$$R^{12}_{12}=\sum_{i+j=k+1}\left(z^1_i z^2_j+z^1_j z^2_i\right),$$
which can be put in the canonical form \eqref{exp} after a linear chenge of basis of the symmetries. The expressions of these expansions in the Rieman invariants can be found by using formulae given in Example \ref{exzak1}.



\subsection{Dispersionless Boussinesq reduction}
The case of the dispersionless Boussinesq reduction can be treated in a similar way. From Example \ref{dbouss}, we will consider a function $\lm$ which is polynomial in $p$, 
$$\lm=p^3+3A^0p+3A^1,$$
thus meromorphic on the Riemann sphere.  The choice of a different normalisation reflects in the expansions below, where we have to consider an expansion in the local parameter $t=\lm^{-\frac{1}{3}}$. For simplicity let us consider only the
  case $k\ge 0$. We observe that, apart from the poles at
$p=v_1$ and $p=v_2$, we have only
  an additional pole at infinity (starting from $k=2$).
Following the same procedure used in the Zakharov case
  we obtain
\begin{align*}
&R^{ij}_{ij}=0,&k=0,1,\\
&R^{ij}_{ij}=\,3\,\,\underset{t=0}{\rm{res}}\left(w^1(t)w^2(t)\,t^{-(3k+4)}\,dt\right), &k>2.
\end{align*}
The expansions of $w^1(t)$ and $w^2(t)$ near $t=0$ are given by
\begin{align*}
w^1(t)=\sum_{k=0}^{\infty}\,k\,w^1_k\,\,t^{k+4}=&\,\, t^4+2\,(-A_0)^{\f{1}{2}}\,t^5
+4\,\left(A_1-(-A_0)^{\f{3}{2}}\right)t^{7}\\
&+5\,\left(2A_1(-A_0)^{\f{1}{2}}-A_0^2\right)t^{8}+\dots\\
&\\
w^2(t)=\sum_{k=0}^{\infty}\,k\,w^2_k\,\,t^{k+4}=&\,\,t^4-2\,(-A_0)^{\f{1}{2}}\,t^5
+4\,\left(A_1+(-A_0)^{\f{3}{2}}\right)t^{7}\\
&+5\,\left(-2A_1(-A_0)^{\f{1}{2}}-A_0^2\right)t^{8}+\dots
\end{align*}
 From these formulas we immediately get the quadratic
expansion of the Riemann tensor:
\begin{eqnarray*}
k=0:&&R^{12}_{12}=0,\\
k=1:&&R^{12}_{12}=0,\\
k=2:&&R^{12}_{12}=3(v^1+v^2)=0,\\
\end{eqnarray*}
More generally, we have
$$R^{12}_{12}=\frac{3}{2}\sum_{i+j=k-1}(w^1_{3i}w^2_{3j}+w^1_{3j}w^2_{3i})\qquad k>2.$$
The expression in the Riemann invariants can be obtained from Example \ref{dbouss}.

\section*{Acknowledgments}
We would like to thank the ESF grant MISGAM 1414, for its support of Paolo Lorenzoni's visit to Imperial.
We are also grateful to the European Commission's FP6 programme for support of this work through the ENIGMA network,
and particularly for their support of Andrea Raimondo, who also received an EPSRC DTA. We would like to thank Maxim Pavlov for valuable discussions, and the integrable system groups at Milano Bicocca and Loughborough universities for their hospitality to Andrea Raimondo.

\bibliographystyle{plain}
\bibliography{mindi}

\end{document}